\documentclass[10pt,letterpaper]{article}

\usepackage{graphicx, amssymb, bm, eucal, float}
\usepackage[square,sort,comma,numbers]{natbib}

\def\a{\alpha}
\def\b{\beta}

\def\e{\epsilon}
\def\f{\phi} \def\vf{\varphi}
\def\g{\gamma}

\def\l{\lambda}
\def\m{\mu}
\def\n{\nu}

\def\r{\rho}
\def\s{\sigma}

\def\L{\Lambda}

\def\fr{\frac}

\def\pro{\propto}
\def\la{\left}
\def\ra{\right}

\def\inf{\infty}

\def\abs#1{\left| #1\right|}

\def\bar#1{\overline{#1}}

\def\ba{\begin{array}}
\def\ea{\end{array}}
\def\be{\begin{equation}}
\def\ee{\end{equation}}
\def\bdm{\begin{displaymath}}
\def\edm{\end{displaymath}}
\def\bea{\begin{eqnarray}}
\def\eea{\end{eqnarray}}
\def\nl{\nonumber \\}

\def\lb{\label}
\def\sp{~~~}

\begin{document}

\title{Solvable potentials in a FLRW+Scalar universe and\\ Fits to type Ia supernovae data}
\author{B. S. Balakrishna\footnote{Email: balakbs2@gmail.com}}
\date{October 25, 2022 \\ Revised: April 14, 2023}

\maketitle

\begin{abstract}

FLRW equations are analyzed in a universe with a cosmic scalar background that is spatially uniform but time-varying. Some solvable potentials to the combined dynamics in such a universe are presented, that are consistent with the scalar dynamics as a consequence of energy momentum conservation. Certain potentials are found to provide very good fits to type Ia supernovae data, with the kinetic and potential energies of the scalar providing the source for dark matter and dark energy. The scalar rolls down the potential as the universe expands with the potential playing the role of a time-varying cosmological constant, modeling a scenario recently discussed in the literature.

\end{abstract}

\section{Introduction}

Type Ia supernovae (SNe Ia) data by the supernova cosmology project team\citep{pxx} has provided us with a description of the contents of the universe. Within the framework of the $\L$CDM model of the universe, the data indicates about 34\% matter and 66\% dark energy. Because observations reveal only about 5\% ordinary matter, the rest 29\% has been attributed to dark matter, leading to dark matter searches. Given the absence of any observational evidence for dark matter, various candidates have been suggested, in particular the possibility that a scalar background could be the underlying source of dark matter/dark energy.

Here such a possibility is further explored. An equivalent set of equations governing a FLRW universe are first set up in a spatially uniform but time-varying scalar background. The equations allow us to construct solvable scalar potentials consistent with the combined dynamics in such a universe, with the equation of motion for the scalar being satisfied as a consequence of energy momentum conservation. Certain potentials are found to provide very good fits to SNe Ia data. An explicit expression is obtained for the potential function $V(\f)$ for the scalar field $\f$ fitting the data exactly as in $\L$CDM:
\be
V(\f) = \a_0+\fr{3\a_0}{8\g^2}{\rm sinh}^2(\g\f), \sp \g=\sqrt{\fr{3(\a_1+\r_1)}{4\a_1}}, ~ \a_0=0.66\r_c, ~ \a_1+\r_1=0.34\r_c, ~ 0\leq\r_1<0.34\r_c,
\ee
where $\r_1$ is the current matter density and $\r_c$ the critical density. If $\r_1$ accounts for just the observed matter density, the rest of the critical density can be thought of as being accounted for by the kinetic and potential energies of the scalar.

In the above model, the scalar $\f$ starts off infinitely large at early times and rolls down the potential to zero as the universe expands. The potential, starting off infinitely large at early times drops asymptotically to a fixed value, the cosmological constant term in the potential. We may say that the scalar provides a source for just the dark matter, but SNe Ia data can be fit equally well with potentials approaching zero as the universe expands. An example of a such potential providing the source for just the dark energy, fitting the data as in a flat-wCDM model, is
\be
V(\f) = \a\la(1-\fr{\b}{6}\ra)\la[\la(\fr{\r_1}{\a}\ra){\rm sinh}^2(\g\f)\ra]^{-\sqrt{\b}/(2\g)}, \sp \g = \fr{3-\b}{2\sqrt{\b}}, ~ \a = 0.73\r_c, ~ \r_1 = 0.27\r_c, ~ \b = 0.5.
\ee
We may then say that the potential played the role of a time-varying cosmological constant, in line with such considerations in the literature under the name of quintessence\citep{rpx,cds,dxx}. An essential part of such a modeling exercise is to propose a viable scalar potential that can account for the universe contents. Here in the article, such potentials are presented in a solvable FLRW+scalar formulation compatible with empirical data. For earlier work on implying scalar potentials from observations, see \citep{htx,ncx,srs,mbs,wax}.

The article is organized as follows. Section (\ref{frw}) sets up the framework involving FLRW equations. Section (\ref{sdx}) discusses its consistency with scalar dynamics. Section (\ref{spx}) presents a class of solvable scalar potentials to the equations. Section (\ref{sne}) provides fits to SNe Ia data. Section (\ref{cls}) concludes with some remarks. Appendix (\ref{pqs}) discusses tractability in a generic context.

\section{FLRW Equations}
\lb{frw}

Let us consider a homogeneous and isotropic universe in the presence of scalar fields $\f_i, i=1,\cdots$, and normal matter referred to as `matter' with no qualifier. Scalar fields are taken to be spatially uniform but time-varying. Let $U$ be the kinetic part of their dynamics under a collective scalar potential $V$. Matter density is $\r$ and its pressure $p$. In this universe with uniform spatial curvature $k$, the FLRW equations for the scale factor $a(t)$ read (in units $c=1$ and $8\pi G=1$)
\bea
\fr{3\dot{a}^2}{a^2}+\fr{3k}{a^2} = &=& U+V+\r, \nl
\fr{2\ddot{a}}{a}+\fr{\dot{a}^2}{a^2}+\fr{k}{a^2} = &=& V-U-p.
\lb{aeq}
\eea
Scale factor is chosen to be unity at present time. A dot on a symbol denotes time differentiation. A prime on a symbol will denote differentiation with respect to $a$. $U$ represents
\be
U = \fr{1}{2}\sum_i\dot{\f_i}^2 = \fr{1}{2}\fr{\dot{a}^2}{a^2}W, \sp W=\sum_ia^2\f_i'^2,
\ee
which introduces $W$ for convenience. After expressing $U$ in terms of $W$ in the first of Eqs. (\ref{aeq}) and rearranging to get the $\dot{a}^2$'s together, we get
\be
\fr{3\dot{a}^2}{a^2} = \fr{V+\r-3k/a^2}{1-W/6}.
\ee
Another relation follows by adding the two Eqs. in (\ref{aeq}), multiplying by $a^5$ and integrating,
\be
\fr{3\dot{a}^2}{a^2} = \fr{C_{\e}}{a^6}-\fr{3k}{a^2}+\fr{3}{a^6}\int_{\e}^adaa^2\la(2a^3V+(\r-p)a^3\ra),
\lb{bcv}
\ee
where $C_{\e}$ is an integration constant and $\e$ is a suitable lower bound on $a$.

For our purpose, it is convenient to express the equations in terms of the scalar energy density $f(a)$, that we may call a fit-function for use in our SNe Ia data analysis, so that
\be
\fr{3\dot{a}^2}{a^2} = f(a)+\r-\fr{3k}{a^2}, \sp f(a) = U+V.
\ee
$f(a)$ would satisfy $f(1)=\r_c-\r_1+3k$ where $\r_c$ is the current critical density and $\r_1$ is the current matter density. In terms of $f$, we have
\bea
f(a)+\r-\fr{3k}{a^2} &=& \fr{V+\r-3k/a^2}{1-W/6}, \nl
f(a)+\r-\fr{3k}{a^2} &=& \fr{C_{\e}}{a^6}-\fr{3k}{a^2}+\fr{3}{a^6}\int_{\e}^adaa^2\la(2a^3V+(\r-p)a^3\ra).
\eea
This lets us express both $V$ and $W$ in terms of $f$. Differentiating the second equation gives us $V$ that can be used in the first to obtain $W$, so that
\bea
V &=& f+\fr{1}{6}\la(af'+a\r'+3(\r+p)\ra), \nl
W &=& -\fr{1}{\la(f+\r-3k/a^2\ra)}\la(af'+a\r'+3(\r+p)\ra).
\lb{vf1}
\eea
Combination $a\r'+3(\r+p)$ appearing above stands for a sum of such combinations for each kind of matter. In an adiabatic evolution of the universe, in the absence of matter creation, each such combination will equate to zero as a consequence of the respective fluid equation, so that we have
\be
a\r'+3(\r+p) = 0.
\lb{feq}
\ee
This simplifies the set of equations further to, assuming a single scalar background and a spatially flat universe with $k=0$,
\be
V = f+\fr{a}{6}f', \sp a^2\f'^2 = \fr{-af'}{f+\r}.
\lb{vf2}
\ee
Note that for $U$ we just have $U=f-V=-af'/6$. Above results hence indicate a split up of the fit-function between kinetic $U$ and potential $V$ components of the scalar.

Results for $V$ and $\f$ provide a relation between them in terms of `parameter' $a$. In other words, it could be viewed as a parametric description of the potential $V(\f)$ given a fit-function $f(a)$. To obtain $V(\f)$ as a function of $\f$ itself, one could attempt to solve for $a$ in terms of $\f$ from the second equation and use it in the first. As we will see in section (\ref{sdx}), such solvable potentials provide a self-consistent framework, with the equation of motion of $\f$ being automatically satisfied. Explicitly, for $\f(a)$ we have
\be
\f(b) = \pm\int^b\fr{da}{a}\sqrt{\fr{-af'}{f+\r}}.
\lb{peq}
\ee
Base limit of integration could be taken to be 0 if $\f\to 0$ as $a\to 0$, or $\inf$ if $\f\to 0$ as $a\to\inf$, or something in-between if $\f$ diverges on both ends.

Fluid equation for matter can be handled using co-moving matter densities $\s$ for each kind of matter, given its equation of state $p=w\r$:
\be
\r = \s a^{-3(1+\bar{w})}, \sp \bar{w} = \fr{1}{{\rm ln}a}\int_1^a\fr{db}{b}w(b).
\lb{six}
\ee
For constant $w$, we have $\bar{w}=w$. In terms of $\s$, fluid equation simply reads $\s'=0$ for each kind of matter. It is helpful to analyze the scalar dynamics as well in terms of its equation of state parameter $w_s$:
\be
w_s = \fr{U-V}{U+V} = -1-\fr{a}{3f}f'-\fr{1}{3f}\la(a\r'+3(\r+p)\ra).
\lb{eqs}
\ee
A useful relation between $w_s$ and $a\f'$ is, assuming Eqs. (\ref{feq}) and (\ref{vf2}),
\be
w_s = -1-\fr{a}{3f}f' = -1+\fr{1}{3f}(f+\r)a^2\f'^2 ~\geq~ -1+\fr{1}{3}a^2\f'^2,
\lb{wse}
\ee
where the equality holds in the absence of matter. The range of values $w_s$ takes is hence related to the rolling speed $\abs{a\f'}$ the scalar field takes as it runs through the potential.

\section{Scalar Dynamics}
\lb{sdx}

Its interesting and important to note that, in a single scalar background $\f$ with the fluid equation (\ref{feq}) assumed to hold, potential $V$ is automatically consistent with the equation of motion for $\f$ as a consequence of conservation of energy momentum, or rather we could say it is implied to be consistent with the equation of motion for $\f$. To see this, let us use Eq. (\ref{bcv}) in the first of Eqs. (\ref{aeq}),
\be
U = \fr{C_{\e}}{a^6}-\fr{1}{a^6}\int_{\e}^adaa^3\la(a^3V'+(\r a^3)'+3pa^2\ra),
\lb{Ueq}
\ee
leading to
\be
(a^6U)'+a^6V'+a^3(\r a^3)'+3pa^5 = 0.
\lb{dva}
\ee
One can check explicitly that our expressions for $U$ and $V$ do satisfy this relation. Now, using $U=\dot{\f}^2/2$, multiplying by $\dot{a}$ and going from $a-$differentiation to $t-$differentiation, we get
\be
\fr{1}{2}\fr{d}{dt}\la(a^3\fr{d\f}{dt}\ra)^2+a^6\fr{dV}{dt}+a^3\fr{d(\r a^3)}{dt}+3pa^5\fr{da}{dt} = 0.
\ee
Rewriting using $d/dt=(d\f/dt)d/d\f$ in the second term and dividing by $a^6$ gives
\be
\fr{d\f}{dt}\la[\fr{1}{a^3}\fr{d}{dt}\la(a^3\fr{d\f}{dt}\ra)+\fr{dV}{d\f}\ra]+\fr{1}{a}\fr{da}{dt}\la[a\r'+3(\r+p)\ra] = 0.
\lb{fre}
\ee
With the fluid equation (\ref{feq}) satisfied, the second term vanishes, and we obtain the equation of motion for the scalar in the $a-$background:
\be
\fr{1}{a^3}\fr{d}{dt}\la(a^3\fr{d\f}{dt}\ra)+\fr{dV}{d\f} = 0.
\ee
Thus, working with a fit-function is consistent with the scalar dynamics for its implied potential $V$.

If there are any couplings of scalar to matter, they can make their appearance along the lines of $\r$ and $p$ in our expressions. Let us assume that they couple collectively to $\r$ as $I_{\r}\r$ and to $p$ as $I_pp$, with some field-dependent interaction strengths $I_{\r}$ and $I_p$ respectively. Using this in Eq. (\ref{dva}) along the lines of $\r$ and $p$, with the time-derivatives of $I_{\r}$ and $I_p$ consumed by the equations of motion, gives us
\be
\la(1+I_{\r}\ra)\la(\r a^3\ra)'+3\la(1+I_p\ra)pa^2 = 0.
\lb{mcx}
\ee
Assuming just one kind of matter with equation of state parameter $w$, we can rewrite this for its co-moving density $\s$ (introduced in Eq. (\ref{six})) as
\bea
a\s' &=& 3w\fr{(I_{\r}-I_p)}{(1+I_{\r})}\s, \sp \s = \r a^{3(1+\bar{w})}, \nl
\s(a) &=& \s(\e){\rm exp}\la(3\int_{\e}^a\fr{da}{a}w(a)\fr{(I_{\r}-I_p)}{(1+I_{\r})}\ra),
\lb{mcn}
\eea
where $\e$ is a suitable lower cutoff for $a$. For specifics let us consider just one scalar, coupling linearly as $I_{\r}=\l_{\r}\f>0$ and $I_p=\l_p\f>0$. If $\l_{\r}>\l_p>0$ and $w\neq 0$, this would result in matter continuously being created. Any initial perturbation in the energy density can seed further creation. This is of course too simplistic a view of matter creation. It doesn't specify the contents or mechanism, a Boltzmann equation arising as a consequence of energy momentum conservation. However, it is interesting to note that it's rate is proportional to equation of state parameter $w$ (assuming $I_{\r}$ and $I_p$ are independent of $w$), suggesting that matter creation was active early on in the history of the universe and is largely suppressed at later times.

If instead interested in expressing the above in terms of couplings $I_u$ and $I_v$ respectively to the 'kinetic' and 'potential' components of matter, one could either re-derive it along the lines of $U$ and $V$ in Eq. (\ref{dva}), or replace $I_{\r}$ and $I_{p}$ in the above result with (for constant $w$)
\bea
I_{\r} &=& \fr{1}{2}\la[(I_u+I_v)+w(I_u-I_v)\ra], \nl
I_p &=& \fr{1}{2w}\la[(I_u-I_v)+w(I_u+I_v)\ra].
\eea
The contribution of matter couplings to $V$ in our Eqs. (\ref{vf1}) would be an addition of
\be
\fr{a}{6}\f'(a)\fr{dI_{\r}(\f)}{d\f}\r,
\ee
and its negative to $\sum a^2\f'^2$ inside parenthesis (as well as $I_{\r}\r$ in the denominator). There are other contributions involving $I_\r$ and $I_p$, but those will get consumed if we require that Eq. (\ref{mcx}) is satisfied.

\section{Solvable Potentials}
\lb{spx}

Let us consider a flat universe ($k=0$) with co-moving matter densities constant in time, unless specified otherwise. Let us build models for some chosen fit-functions, since working with fit-functions directly helps us to be in better agreement with SNe Ia data.

There are some generic characteristics that are helpful in choosing fit-functions. For $U$ to remain non-negative, $f$ need to be a non-increasing function of $a$. Also, for $V$ to be bounded from below by say $V_L$, $a^6(f-V_L)$ need to be a non-decreasing function of $a$. In other words, $f$ would be non-increasing, and if decreasing, doing so not faster than $1/a^6$.

To start with, let us consider a fit-function that mimics that of $\L$CDM:
\be
f(a) = \a_0+\fr{\a_1}{a^3}.
\lb{f03}
\ee
This is solvable in the presence of matter density $\r=\r_1/a^3$ ($\r_1$ being the current matter density):
\bea
\f(a) &=& \fr{1}{\g}~{\rm ln}\la(\sqrt{\b}a^{-3/2}+\sqrt{1+\b a^{-3}}\ra), \sp \b=\fr{\a_1}{\a_0}, \nl
V(\f) &=& \a_0+\fr{\a_1}{2a^3} = \a_0+\fr{3\a_0}{8\g^2}{\rm sinh}^2(\g\f), \sp \g=\sqrt{\fr{3(\a_1+\r_1)}{4\a_1}}.
\lb{f13}
\eea
Another interesting choice is the one mimicking that of flat-wCDM:
\be
f(a) = \fr{\a_0}{a^{\b}}+\fr{\a_1}{a^3}, \sp 0<\b<3.
\ee
However, it is not explicitly solvable in general, but for the case where all of matter density is attributed to $\r_1$, that is when $\a_1=0$, it implies
\bea
\f(a) &=& \fr{1}{\g}{\rm arsinh}\la(\sqrt{\fr{\a_0}{\r_1}}a^{(3-\b)/2}\ra), \sp \nl
V(\f) &=& \a_0\la(1-\fr{\b}{6}\ra)\la[\la(\fr{\r_1}{\a_0}\ra){\rm sinh}^2(\g\f)\ra]^{-\sqrt{\b}/(2\g)}, \sp \g = \fr{3-\b}{2\sqrt{\b}}.
\eea
These models are discussed in the context of a data fit to SNe Ia in section (\ref{sne}). 

Choice (\ref{f03}) belongs to a class of fit functions of the form
\be
f(a) = \fr{1}{\la(\a_0+\a_1a^{\n}\ra)^n}, \sp 0<\n n\leq 6,
\ee
where $\n$ and $n$ are both together positive or together negative. They can be analyzed parametrically in the presence of matter. They are solvable in the absence of matter giving
\bea
\f(a) &=& \fr{1}{\g}~{\rm ln}\la(\sqrt{\b}a^{\n/2}+\sqrt{1+\b a^{\n}}\ra), \sp \b=\fr{\a_1}{\a_0}, \nl
V(\f) &=& \fr{\a_0+(1-\n n/6)\a_1a^{\n}}{\la(\a_0+\a_1a^{\n}\ra)^{n+1}}, \sp a^{\n}=\fr{1}{\b}{\rm sinh}^2\la(\g\f\ra), \sp \g=\sqrt{\fr{\n}{4n}}.
\lb{fnn}
\eea
Integration limit in the $\f$ expression is taken to be zero when $\n,n$ are positive and $\inf$ when $\n,n$ are negative, with the sign chosen appropriately to keep $\f$ positive. The potential is even in $\f$, but the two cases have opposite behavior. When $\n,n$ are negative, the potential is an increasing function of $\f$, having at a finite value at $\f=0$ and rising as $\f\to\inf$. As for $\f$, it starts off from a infinitely large value at time zero and rolls down the potential, approaching zero as the universe expands. The potential too starts off infinitely large but approaches a constant as $a\to\inf$, playing  the role of a cosmological constant asymptotically. On the other hand, when $\n,n$ are positive, the potential is a decreasing function of $\f$, has a finite value at $\f=0$ and drops to zero as $\f\to\inf$. In this case, $\f$ starts off from zero at time zero and rolls down the potential, running away to $\inf$ as universe expands. The potential starting off at a finite value approaches zero as $a\to\inf$. We may say that the potential played the role of a time-varying cosmological constant or that the cosmological constant lived for a short period during the beginning of time. Such a potential may also have applications in building cosmic inflationary models.

The are other fit-functions solvable in the absence of matter, some of which are likely to be known, like for instance (choosing $\f'>0,~\f(1)=0$)
\bea
f(a) &=& Aa^{-\b}, \sp \f(a) = \sqrt{\b}~{\rm ln}a, \nl
V(\f) &=& A\la(1-\fr{\b}{6}\ra)e^{-\sqrt{\b}\f},
\lb{ef1}
\eea
which yields a potential exponential in $\f$. The following gives it a quadratic dependence in the exponential (choosing $\f\ge 0,~\f(0)=0$),
\bea
f(a) &=& Ae^{-\a a^{\b}}, \sp \f = 2\sqrt{\fr{\a}{\b}}a^{\b/2}, \nl
V(\f) &=& A\la(1-\fr{1}{24}\b^2\f^2\ra)e^{-\b\f^2/4}.
\lb{ef2}
\eea
Here, the potential turns negative for a while, but $f,U$ are positive quantities. Solution with $\a,\b$ negative leads to $V$ unbounded from below. More generally, one can construct solvable potentials from Eq. (\ref{tfb}) by simply choosing a $u(\f)$, or from Eq. (\ref{tfc}) in the absence of matter.

Let us analyze the approach to $a\to 0,\inf$ in some generality in the absence of matter. Consider the case where $f$ tends to a finite nonzero constant $f(0)$ as $a\to 0$. Since $f$ needs to be non-increasing, it may tend to $f(0)$ as say $f\to f(0)\la(1-\a a^{\b}\ra)$ for some positive $\a,\b$. This suggests that $a^2\f'^2\to \a\b a^{\b}$ and $\f\to 2\sqrt{(\a/\b)}a^{\b/2}$ (choosing $\f\ge 0,~\f(0)=0$), implying
\be
V = f+\fr{a}{6}f' \to f(0)-\fr{1}{24}\b(6+\b) f(0)\f^2.
\ee
This is an inverted quadratic potential, at least locally near $\f=0$. Inverted potentials are known to be useful in modeling cosmic inflation. Alternatively, consider the case where $f$ tends to $f(0)$ as $f\to f(0)\la(1-\a(-x)^{-\b}\ra)$ where $x={\rm ln}a$. Then $a^2\f'^2\to \a\b(-x)^{-\b-1}$ and $\f\to \g(-x)^{-(\b-1)/2}$ (assuming $\b\ne 1$, $\g$ defined below). This implies
\be
V \to f \sim f(0)-\a f(0)(\f/\g)^{2\b/(\b-1)}, \sp \g = \fr{2\sqrt{\a\b}}{\b-1}.
\ee
Case $\b=1$ is special with $f\to f(0)\la(1+\a/x\ra)$, so that $a^2\f'^2\to \a/x^2$ and $\f\to-\sqrt{\a}{\rm ln}(-x)$, implying $V\to f(0)-\a f(0)e^{\f/\sqrt{\a}}$. This corresponds to potentials with an exponential tail defining a nearly flat plateau for large negative values of $\f$, also useful in modeling cosmic inflation.

Opposite case of $f$ tending to $f(\inf)>0$ as $a\to\inf$ can be analyzed similarly with $f\to f(\inf)\la(1+\a a^{-\b}\ra)$, $a^2\f'^2\to \a\b a^{-\b}$, $\f\to 2\sqrt{(\a/\b)}a^{-\b/2}$ and
\be
V \to f(\inf)+\fr{1}{24}\b(6-\b) f(\inf)\f^2.
\ee
This is a convex potential, at least locally near $\f=0$. Alternatively, one may have $f\to f(\inf)\la(1+\a x^{-\b}\ra)$, $a^2\f'^2\to \a\b x^{-\b-1}$, $\f\to \g x^{-(\b-1)/2}$ ($\b\ne 1$, $\g$ defined below) and
\be
V \to f \sim f(\inf)+\a f(\inf)(\f/\g)^{2\b/(\b-1)}, \sp \g = \fr{2\sqrt{\a\b}}{\b-1}.
\ee
For the special case $\b=1$ we have $f\to f(\inf)\la(1+\a/x\ra)$, $a^2\f'^2\to \a/x^2$ and $\f\to \sqrt{\a}{\rm ln}(x)$, implying $V\to f(\inf)+\a f(\inf)e^{-\f/\sqrt{\a}}$.

The case of either $f(0)=\inf$ or $f(\inf)=0$ needs to be treated differently. If $f\to \a a^{-\b}$ as $a\to 0$, we have $a^2\f'^2\to\b$, $\f\to \sqrt{\b}{\rm ln}a$ (choosing $\f'>0,~\f(1)=0$), and hence
\be
V = f+\fr{a}{6}f' \to \a\la(1-\fr{\b}{6}\ra)e^{-\sqrt{\b}\f}.
\ee
These limits hold for $a\to\inf$ as well. They are consistent with the above solved model with $\b=\n n$. Alternatively, if $f\to\a(-x)^{\b}$ as $a\to 0$ where $x={\rm ln}a$, we have $a^2\f'^2\to-\b/x$, $\f\to-\g\sqrt{-x}$ and $V\to \a(-\f/\g)^{2\b}$ where $\g=2\sqrt{\b}$. For $a\to\inf$, $f\to\a x^{-\b}$, $a^2\f'^2\to\b/x$, $\f\to\g\sqrt{x}$ and $V\to \a(\f/\g)^{-2\b}$.

\section{Fit to SNe Ia data}
\lb{sne}

Let us again consider a flat universe ($k=0$) and pressure-less matter ($w=0$). Given a fit-function $f(a)$, one obtains the proper distance light travels as a function of red-shift $z$ as,
\be
d(z) = \fr{c}{H_0}\int_{1/(1+z)}^1\fr{da}{a^2}\sqrt{\fr{f(1)+\r_1}{f(a)+\r}},
\ee
where $H_0$ is the current value of the Hubble constant taken to be 73km/sec/Mpc, $\r=\r_1/a^3$ is the co-moving matter density and $\r_1$ is the current matter density. Using this in the following formula generates the distance-modulus vs red-shift curve as a fit to SNe Ia data:
\bea
\m(z) = 5{\rm log}_{10}((1+z)d(z))+25, \sp {\rm where~d~is~in~Mpc}.
\eea
We will choose a certain forms for the fit-functions and explore its consistency with the SNe Ia data.

There are many one could try that fit the data comparably to $\L$CDM. Let us start with the simplest, one that is closest to that fit, that fit itself (model I):
\be
f(a) = \a_0+\fr{\a_1}{a^3}, \sp \a_0 = 0.66\r_c, ~ \a_1+\r_1 = 0.34\r_c,
\lb{mdl1}
\ee
where $\r_c$ is the current critical density. The resulting fit is exactly the one that the $\L$CDM model generates, hence it would fit the data just as well. The fit parameter $\a_1+\r_1=0.34\r_c$ corresponds to 34\% of the critical density that $\L$CDM suggests for the matter density. However, in our present case, we haven't separated the matter density from the scalar contribution yet. Expressions for $\f$ and $V$ are presented earlier in Eqs. (\ref{f13}).

$\L$CDM is the limiting case of this model as $\a_1\to 0$, or equivalently as $\r_1\to 0.34\r_c$. In our case, the fit can accommodate any matter density up to 34\%. If we knew what the actual scalar potential should be, this would provide us with a prediction for the matter density. Here we simply choose it to be say 5\% of the critical density to be close to observations. The remaining 29\% is supplied by the kinetic and potential energies of the scalar. However, the 66\% that $\L$CDM attributes to the cosmological constant is bundled into the scalar potential. This is a special model in which the dark matter component (29\% at present times) is equally split between the kinetic and the 'true' potential energies at all times, effectively having zero pressure and contributing as cold dark matter. 

An alternative without a cosmological constant like term is to replicate flat-wCDM (model II):
\be
f(a) = \fr{\a_0}{a^{\b}}+\fr{\a_1}{a^3}, \sp \a_0 = 0.70\r_c, ~ \a_1+\r_1 = 0.30\r_c, ~ \b=0.5.
\lb{mdl2}
\ee
The model is solvable when $\a_1=0$, that is when all of matter density is attributed to $\r_1$ so that the model just provides for dark energy with equation of state parameter $\b/3-1=-0.83$. The behavior turns out to be similar except that the scalar potential approaches zero asymptotically as the universe expands. A model with similar behavior is obtained, also without a cosmological constant like term, using (model III)
\be
f(a) = \fr{\a_1}{a}+\fr{\a_2}{a^2}+\fr{\a_3}{a^3}, \sp \a_1 = 1.02\r_c, ~ \a_2 = -0.31\r_c, ~ \a_3+\r_1 = 0.29\r_c.
\lb{mdl3}
\ee
This is positive and a decreasing function of $a$ as required, and provides an equally good fit to data. Similar fit is obtained with $f=0.95\r_c/(a^3(1-2.25{\rm ln}a))$.

There are other alternatives, providing reasonable fits to data but with different potential characteristics. Though the parametric expressions are not easy to solve in the presence of matter, they do admit exact solutions in the absence of matter providing good insights into their behavior. An interesting case is that of Eqs. (\ref{fnn}) with $\n=1,n=3$ (model IV):
\be
f(a) = \fr{1}{\la(\a_0+\a_1a\ra)^3}, \sp \a_0=0.42, ~ \a_1=0.63, ~ \r_1=0.05.
\lb{mdl4}
\ee
This provides a comparable fit to Sne Ia data, but the implied potential has opposite behavior compared to the ones discussed above. Also, as noted below, these models will likely fail to be consistent with CMB data. Such potentials may play a role in the inflationary phase of the scalar field.

Results are presented in Figs. (1-4). Model fits to SNe Ia data are presented in Fig. (1). Evolutions of the scalar potentials with respect to the scale factor are shown in Fig. (2). Deceleration parameters $q$ are plotted against the scale factor in Fig. (3), where $q$ is defined as
\be
q = -\fr{a\ddot{a}}{\dot{a}^2}=-\fr{(a^2(f+\r))'}{2a(f+\r)}.
\ee
Proper distances to current time in units of $c/H_0$ are plotted against the scale factor in Fig. (4). Near closeness of the model's proper distance to that of the $\L$CDM at early times would help the model to be consistent with CMB data. Model IV would fail in this regard. Interestingly, models II and III indicate slightly higher proper distances as compared to $\L$CDM, that could be brought in line with that of the later by rising $H_0$. This can have favorable implications for $H_0$ consistency between CMB and SNe Ia data. However, since the proper distance is calculable directly off the fit-function itself, it is not dependent on our assumption that it is being driven by a scalar field.

\section{Conclusions}
\lb{cls}

As noted in the literature and explored further here, a scalar field can be conveniently modeled to source dark matter/dark energy in the universe today. It can be formulated analytically as a spatially uniform but a time-varying background within the context of a FLRW universe consistent with its scalar dynamics. As shown, there exist solvable scalar potentials with some providing very good fits to SNe Ia data, with the dark matter/dark energy supplied by the kinetic and potential energies of the scalar. As presented elsewhere, a numerical framework can also be set up to investigate the implications of generic scalar potentials. Within the framework of a $\L$CDM model of the universe, SNe Ia data indicates the presence of dark energy in the form of a cosmological constant $\L$. There are solvable models where, besides sourcing dark matter, the scalar potential takes the role of a `time-varying' $\L$ with its potential energy providing the source for dark energy. The dark energy term can appear as bundled into the scalar potential in certain models that can be viewed as the cosmological constant in disguise. But there are also models where the cosmological constant can be considered absent or rather inherently built into the scalar potential.

As is well-known, a concerning thing about the cosmological constant in the $\Lambda$CDM model as a source of dark energy is that it is of the same order of magnitude as the current matter density $\r$. A comparable constant is co-moving matter density $\r a^3$, but that is not invariant under scaling of the spatial coordinates, rather $\r$ is. Since $\r$ is very much a time varying quantity, it is difficult to comprehend why a constant such as $\L$ would be comparable to its current value. In a scenario where the scalar potential takes the role of a `time-varying' $\Lambda$, its value could be comparable to $\r$ at earlier times, perhaps remaining so all the way back until the beginning of the universe. As the data fits indicate, this has an additional advantage of the scalar field capable of providing the source for dark matter. An intriguing follow up then is to identify this scalar field with the inflaton responsible for cosmic inflation in line with such considerations in the literature\citep{sxx, pvx, prx, dvx, wxx}. Models presented here are expected to address the scalar potential back in time only until about last scattering. We could hence view its evolution subsequent to last scattering as being the later part of the inflaton's evolution. Given a model of the inflaton at its earlier times, this could potentially provide a consistent picture of its evolution, and a plausible reasoning for the order of magnitudes of the cosmological parameters, a subject left for further study.

\newpage

\appendix

\section{Potential Equations}
\lb{pqs}

If interested in solving for $f(a)$ given a scalar potential $V(\f)$, one could try to solve the following set of coupled equations (assuming co-moving matter):
\bea
f'+a\f'^2f &=& -\la(\r-3k/a^2\ra)a\f'^2, \nl
V(\f) &=& f+\fr{a}{6}f',
\lb{tfa}
\eea
where, as before, a prime on a symbol denotes differentiation with respect to $a$. These are not tractable in general. One could simplify them by introducing an intermediate function $u(\f)$ defined as
\bea
u(\f(a)) &=& \int^adb~b\f'^2(b), \nl
u_{\f} = \fr{du(\f)}{d\f} &=& a\f'(a), \sp a \pro {\rm exp}\la(\int^{\f}\fr{d\f}{u_{\f}}\ra).
\eea
Integration base limit could be suitably chosen, for instance as zero, one or $\inf$. In terms of $u(\f)$, we can solve the first of Eqs. (\ref{tfa}), given some constant $C$, as
\be
f = Ce^{-u(\f)} + \int_a\fr{db}{b}\la(\r(b)-3k/b^2\ra)u_{\vf}^2e^{u(\vf)-u(\f)},
\lb{tfb}
\ee
where $\f=\f(a)$ and $\vf=\vf(b)$. The potential can now be expressed as
\be
V(\f) = \la(1-\fr{1}{6}u_{\f}^2\ra)f-\fr{1}{6}\la(\r-3k/a^2\ra)u_{\f}^2.
\ee
As for $\r$ and the $k$-term, they can be expressed in terms of $u(\f)$ as
\be
\r-3k/a^2 = \r(p)~{\rm exp}\la(-3\int_{\f(p)}^{\f}\fr{d\f}{u_{\f}}\ra)-(3k/p^2)~{\rm exp}\la(-2\int_{\f(p)}^{\f}\fr{d\f}{u_{\f}}\ra).
\ee
We thus get a nonlinear integro-differential equation for $u(\f)$. Analytic solutions are not available in general, and one needs to resort to numerical computations. The set of results can also be used to imply $V(\f)$ given a choice for $u(\f)$, like for instance as $u(\f)$ quadratic in $\f$. 

Above results become simpler for a flat universe ($k=0$) in the absence of matter ($\r=0$). The second term in the second of Eqs. (\ref{tfb}) is then absent and we just have
\be
f = Ce^{-u(\f)}.
\lb{tfc}
\ee
Using this in the expression for $V(\f)$ gives
\be
V(\f) = C\la(1-\fr{1}{6}u_{\f}^2\ra)e^{-u(\f)}.
\ee
This is a nonlinear first-order differential equation for $u(\f)$, but as before provides for a class of solvable scalar potentials for various choices of $u(\f)$. For instance for the choice $u(\f)\pro \f$ or $\f^2$, we get the result discussed earlier in Eqs. (\ref{ef1}), (\ref{ef2}), or more generally for $u(\f)=\b_0+\b_1+\b_2\f^2/2$, $\f\ge-\b_1/\b_2$. The choice $u-$derivative $u_\f = \sqrt{6}~{\rm tanh}(\b\f)$ gives
\be
V(\f) \pro \la({\rm cosh}(\b\f)\ra)^{-2-\sqrt{6}/\b}, \sp a \pro \la({\rm sinh}(\b\f)\ra)^{1/(\sqrt{6}\b)}.
\ee
This is a special case of (\ref{fnn}) with $\n n=6$. Here, as $a$ runs from $0\to\inf$, the scalar rolls down the potential from $0\to\inf$ for $\b>0$ and from $-\inf\to 0$ for $\b\in\la(0,-\sqrt{6}/2\ra)$. It rolls up the potential from $-\inf\to 0$ for $\b<-\sqrt{6}/2$, and over a constant potential for $\b=-\sqrt{6}/2$.

\newpage

\begin{figure}
\centering
\medskip
\includegraphics[width=4.5in,height=3.5in]{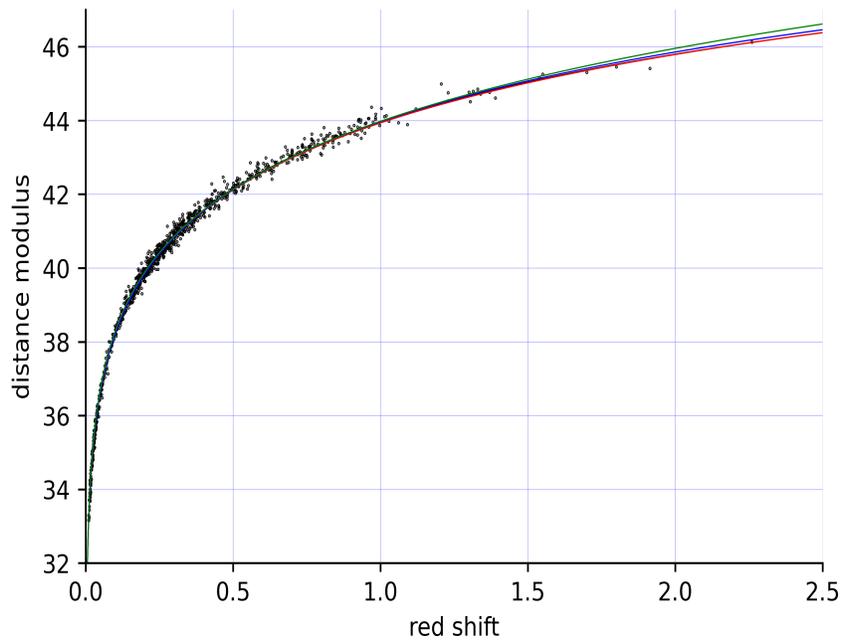}
\caption{Model fits to SNe Ia data. Fit for model of Eq. (\ref{mdl1}) matching $\L$CDM fit is in black. Fits for models of Eqs. (\ref{mdl2}), (\ref{mdl3}) and (\ref{mdl4}) are in red, blue and green respectively.}
\end{figure}

\begin{figure}
\centering
\medskip
\includegraphics[width=4.5in,height=3.5in]{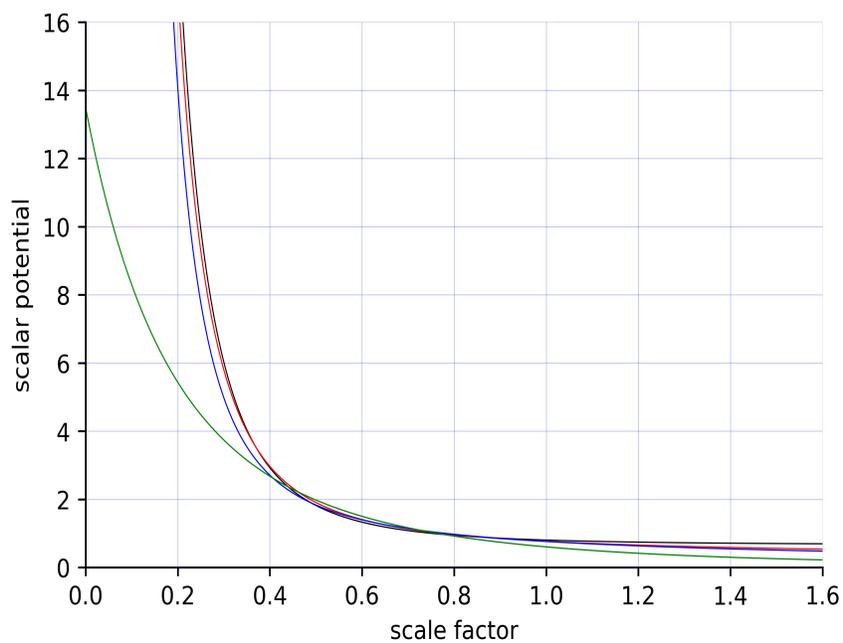}
\caption{Scalar potentials w.r.t scale factor. Potential for model of Eq. (\ref{mdl1}) matching $\L$CDM fit is in black. Potentials for models of Eqs. (\ref{mdl2}), (\ref{mdl3}) and (\ref{mdl4}) are in red, blue and green respectively.}
\end{figure}

\begin{figure}
\centering
\medskip
\includegraphics[width=4.5in,height=3.5in]{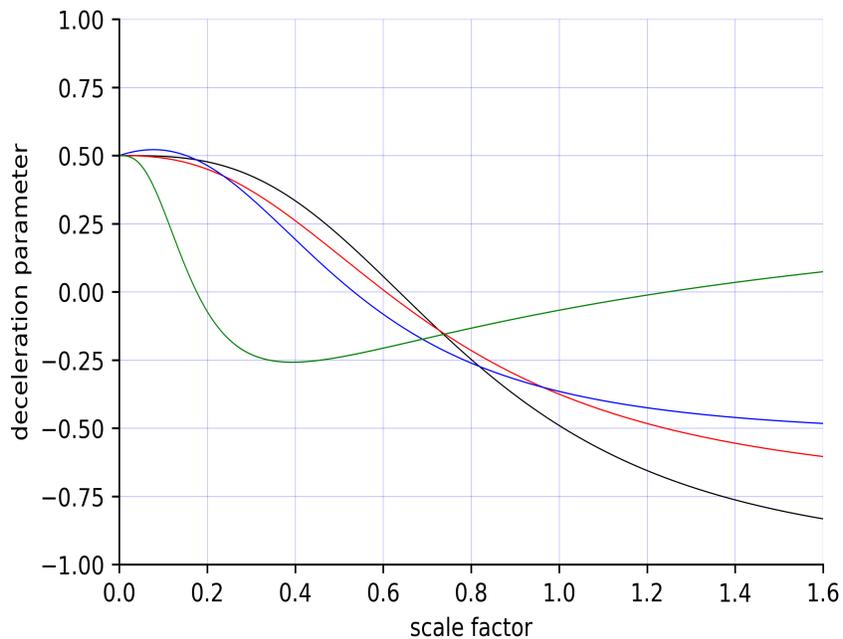}
\caption{Deceleration parameters w.r.t scale factor. Deceleration parameter for model of Eq. (\ref{mdl1}) matching $\L$CDM fit is in black. Deceleration parameters for models of Eqs. (\ref{mdl2}), (\ref{mdl3}) and (\ref{mdl4}) are in red, blue and green respectively.}
\end{figure}

\begin{figure}
\centering
\medskip
\includegraphics[width=4.5in,height=3.5in]{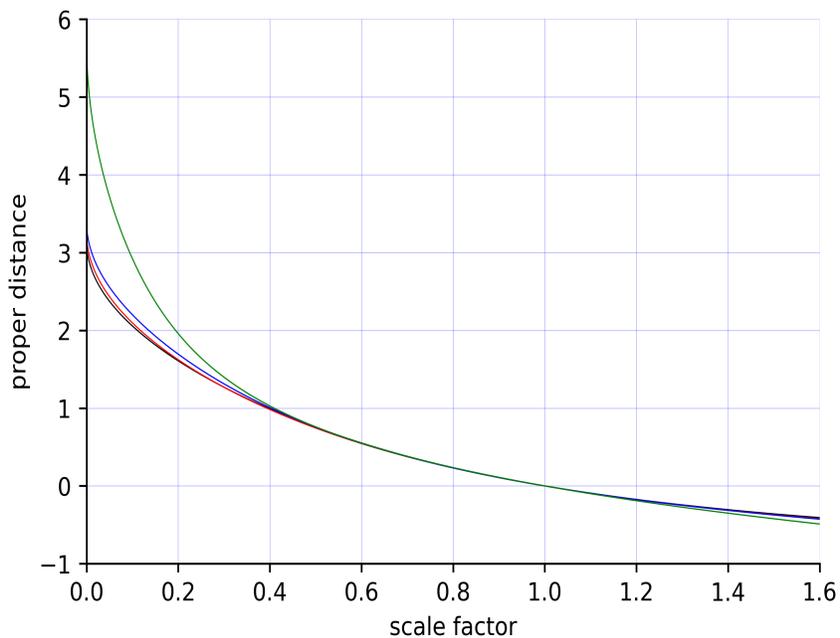}
\caption{Proper distances to current time in units of $c/H_0$ w.r.t scale factor. Proper distance for model of Eq. (\ref{mdl1}) matching $\L$CDM fit is in black. Proper distances for models of Eqs. (\ref{mdl2}), (\ref{mdl3}) and (\ref{mdl4}) are in red, blue and green respectively.}
\end{figure}

\end{document}